\pdfoutput=1

\documentclass[
    ,final            
  ]
  {aipproc}

\layoutstyle{6x9}

\begin{document}
\title{Asymmetric Line Profiles in Spectra of Gaseous Metal Disks Around Single White Dwarfs}
\author{S.~Hartmann}{
  address={Institute for Astronomy and Astrophysics,
    Kepler Center for Astro and Particle Physics,\\
    Eberhard Karls University,
    D-72076 T\"ubingen,
    Germany}}
\author{T.~Nagel}{
  address={Institute for Astronomy and Astrophysics,
    Kepler Center for Astro and Particle Physics,\\
    Eberhard Karls University,
    D-72076 T\"ubingen,
    Germany}}
\author{T.~Rauch}{
  address={Institute for Astronomy and Astrophysics,
    Kepler Center for Astro and Particle Physics,\\
    Eberhard Karls University,
    D-72076 T\"ubingen,
    Germany}}
\author{K.~Werner}{
  address={Institute for Astronomy and Astrophysics,
    Kepler Center for Astro and Particle Physics,\\
    Eberhard Karls University,
    D-72076 T\"ubingen,
    Germany}}
\begin{abstract}
  Around several single DAZ and DBZ white dwarfs metal-rich disks have been observed, which are mostly believed to originate from disruption of smaller rocky planetesimals. In some cases the material does not (only) form a dusty but gaseous disk. In the case of SDSS\,J122859.93+104032.9 the double peaked infrared Ca\,\textsc{ii} triplet $\lambda\lambda$\,8498,\,8542,\,8662\,{\AA}, one of only two emission features of the spectra, exhibits a strong red/violet asymmetry. Assuming a composition similar to a chondrite-like asteroid, being the most prominent type in our own solar system, we calculated the spectrum and vertical structure of the disk using the T\"ubingen non-LTE \emph{Ac}cretion \emph{D}isk \emph{c}ode \textsc{AcDc}. Modified to simulate different non axis-symmetrical disk geometries, the first preliminary results are in good agreement with the observed asymmetric line profile.
\end{abstract}

\maketitle
%
%
\section{Motivation}
\subsection{Metal-Rich Dust Disks}
Starting with the analysis of G\,29-38 by \citet{1987Natur.330..138Z} in 1987, about twenty metal-enriched white dwarfs were discovered, which show a significant infrared excess in their spectra, but for which no cool companions could be found. As these objects have rather short sedimentation times \citep{1997A&A...320L..57K} due to their high surface gravity there has to be an accretion source like a metal-rich dust cloud to explain their high metallicity. In fact, Spitzer observations confirmed the dust material \citep{2005ApJ...635L.161R}, which is located in an equatorial plane \citep{1990ApJ...357..216G} forming a metal-rich but hydrogen- and helium-poor \citep{2003ApJ...584L..91J} dust disk.
\subsection{Metal-Rich Gas Disks}
\begin{figure}%
  \centering%
  \includegraphics[width=0.83\textwidth]{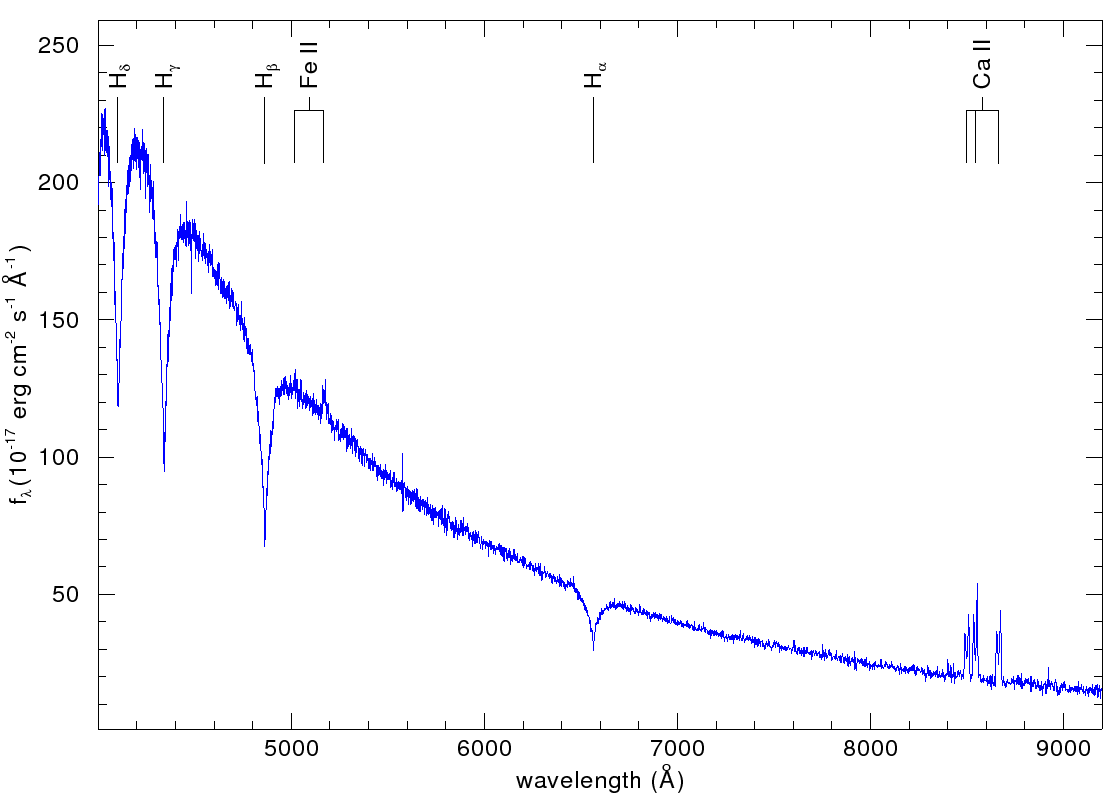}%
  \caption{Spectrum of SDSS\,J1228+1040 from the Sloan Digital Sky Survey.}%
  \label{fig:1}%
\end{figure}

\citet{2006ApJ...646..474K} suggested that the hottest of those white dwarfs, with $T_{\mathrm{eff}}\ge 20\,000\,\mathrm{K}$, might sublimate their dust material and feature a gaseous disk instead. In fact, five white dwarfs have been found in the Sloan Digital Sky Survey (e.g. Fig.\,\ref{fig:1}) by \citet[priv.~comm.]{2006Sci...314.1908G, 2007MNRAS.380L..35G, 2008MNRAS.391L.103G} indicating gas disk emission features in their spectra in addition to the IR-spectral component of a dust disk. These gas features are mainly the Ca\,\textsc{ii} $\lambda\lambda$\,8498,\,8542,\,8662\,{\AA}, although at least two of the five spectra also show a Fe\,\textsc{ii} $\lambda\lambda$\,5018,\,5169\,{\AA} in emission.
\subsection{Origin of the Disk Material}
Theories predict that former planetary systems around the white dwarfs progenitor have a chance to survive the host stars' late evolutionary phases (e.g. \citep{2008AJ....135.1785J}). After the actual formation of the white dwarf, a smaller rocky body like an asteroid or planetesimal might have been disturbed in its orbital motion by another surviving, larger planet and eventually reached the inner part of the system. Close to the central object, tidal forces will disrupt the asteroid providing the material to form the dust and gaseous disk.
\section{Aim and Method}
\subsection{Asymmetric Line Profiles}
\begin{figure}%
  \centering%
  \includegraphics[width=0.55\textwidth]{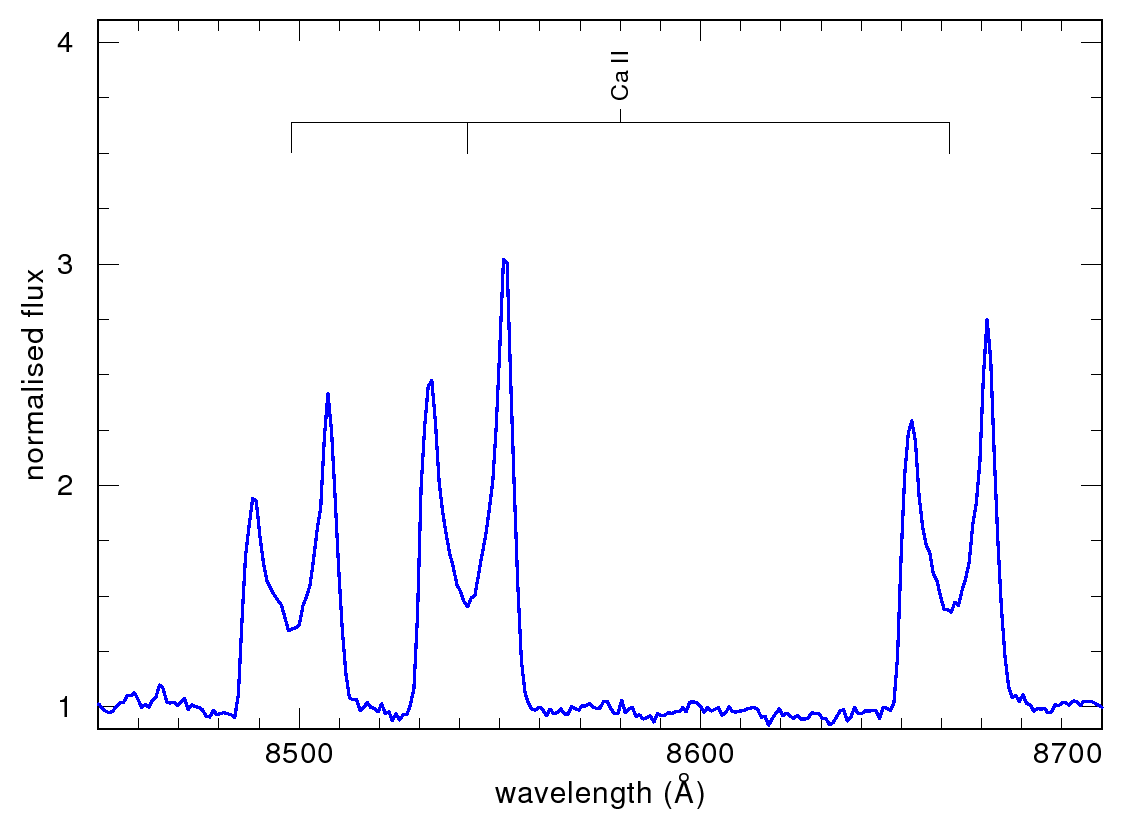}%
  \caption{Ca\,\textsc{ii} $\lambda\lambda$\,8498,\,8542,\,8662\,{\AA} in the spectrum of SDSS\,J1228+1040. Each line is splitted due to the Kepler rotation of the disk but shows an asymmetry in the relative strengths of the red/violet peaks, favoring the red wing peak in this case.}%
  \label{fig:2}%
\end{figure}

Taking a closer look at the prominent Ca\,\textsc{ii} feature of SDSS\,J122859.93+104032.9 (hereafter SDSS\,J1228+1040) (Fig.\,\ref{fig:2}) one recognizes the expected double-peak structure of the lines due to the Kepler rotation of the emitting disk. But unexpectedly, the lines show a significant difference in the red/violet peaks line strengths. We assumed this asymmetry to directly originate from an actual asymmetric disk geometry. To investigate this further we simulated a hydrodynamic disk-evolution scenario to get plausible geometries and later modified our spectrum-simulation code accordingly.
\subsection{Hydrodynamic Simulation with \textsc{fargo}}
\begin{figure}%
  \begin{minipage}{0.45\textwidth}%
    \centering%
    \includegraphics[width=\textwidth]{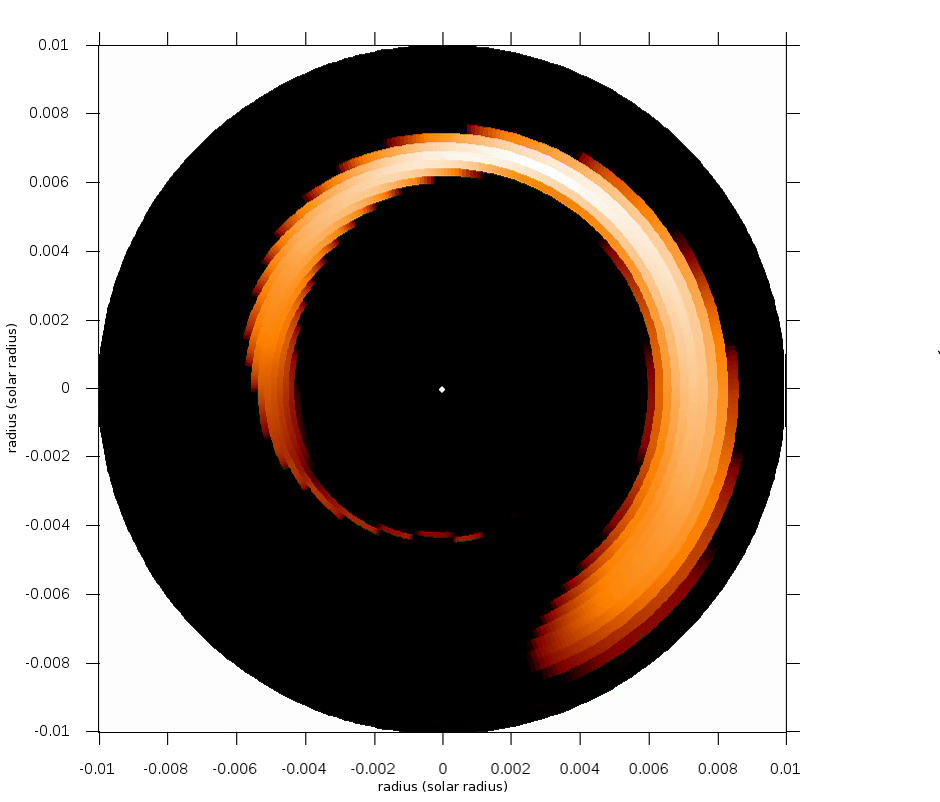}%
    \label{fig:3}%
  \end{minipage}%
  \hfill%
  \begin{minipage}{0.45\textwidth}%
    \centering%
    \includegraphics[width=\textwidth]{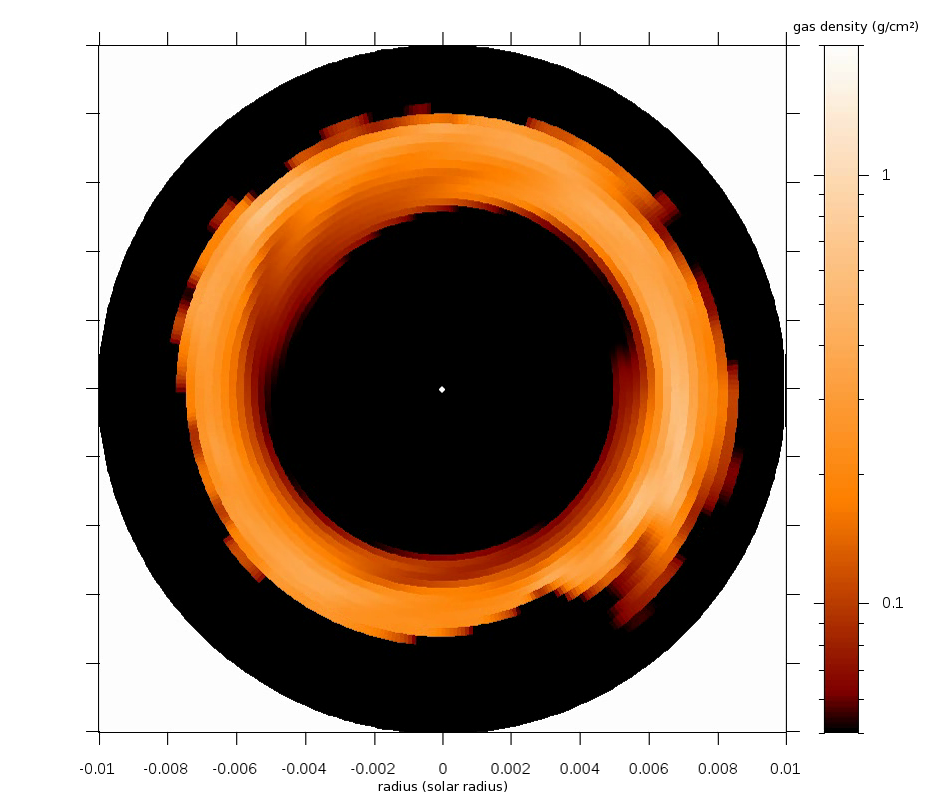}%
    \caption{The two selected evolution phases of the \textsc{fargo} simulation -- color-coded is the surface mass density on a logarithmic scale. The left panel shows an early stage $\left(t_{\mathrm{sim}}\approx 40\,\mathrm{yr}\right)$ as the disks forms a spiral-arm like structure. The right panel shows a late $\left(t_{\mathrm{sim}}\approx 125\,\mathrm{yr}\right)$, but with several hundreds of orbits, long-lasting phase of a fully closed disk with an out-of-balance matter distribution.}%
    \label{fig:4}%
  \end{minipage}%
\end{figure}

To simulate the geometrical evolution of a gaseous disk we used the code \emph{F}ast \emph{A}dvection in \emph{R}otating \emph{G}aseous \emph{O}bjects (\textsc{fargo}) by \citet{2000A&AS..141..165M}. Although this hydrodynamic code is suited for all kinds of sheared fluid disks, it is almost exclusively used in the calculation for planet formation scenarios. We calculated the temporal evolution of a blob of material with Gaussian density distribution (parameters given in Tab.\,\ref{tab:1}).\\
\begin{table}%
  \begin{tabular}{p{8em}lr}\hline
    White dwarf mass&$M_{\mathrm{WD}}/\mathrm{M}_{\odot}$&$0.77$\\
    White dwarf radius&$R_{\mathrm{WD}}/\mathrm{km}$&$7700$\\
    Disk mass&$M_{\mathrm{disk}}/\mathrm{g}$&$7\cdot 10^{21}$\\
    Disk outer radius&$R_{\mathrm{o}}/R_{\mathrm{WD}}$&$136$\\
    Simulation rim&$R_{\mathrm{sim}}/R_{\mathrm{o}}$&$1.5$\\\hline
  \end{tabular}%
  \caption{\textsc{fargo} simulation parameters.}%
  \label{tab:1}%
\end{table}
As a result of this, we selected two geometries, a spiral arm structure of the early simulation phase and a fully closed but out-of-balance disk from the late stage of the evolution (Fig.\,\ref{fig:4}) to perform the modifications on the spectrum synthesis code.
\subsection{Spectral Synthesis Code \textsc{AcDc}} 
\begin{figure}%
  \begin{minipage}{0.45\textwidth}%
    \centering%
    \includegraphics[width=\textwidth]{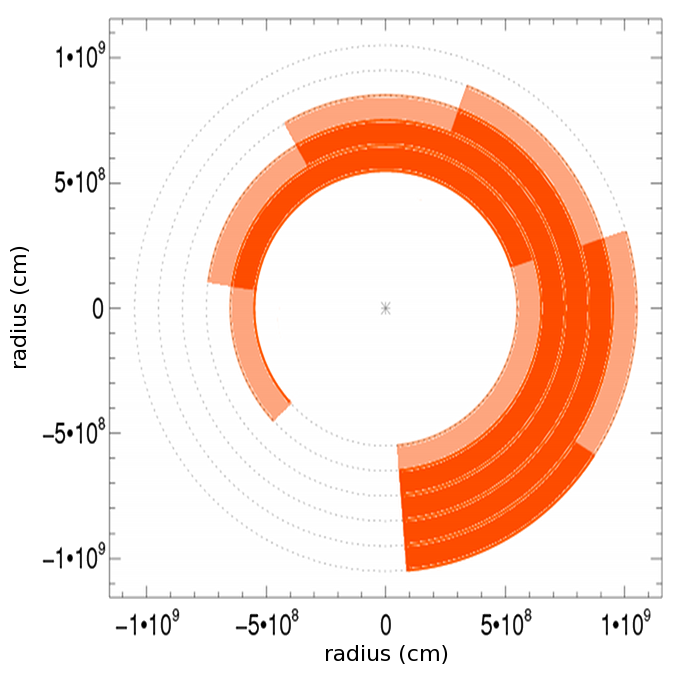}%
    \label{fig:5}%
  \end{minipage}%
  \hfill%
  \begin{minipage}{0.45\textwidth}%
    \centering%
    \includegraphics[width=\textwidth]{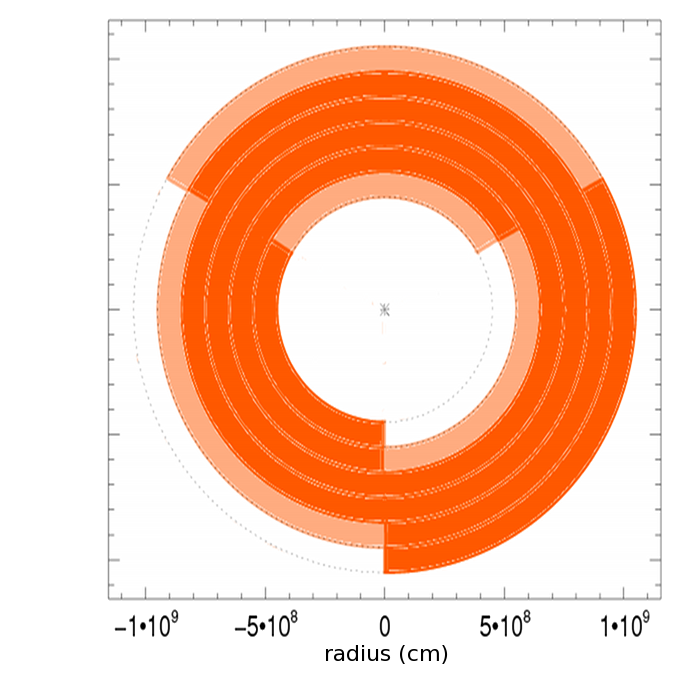}%
    \caption{Geometry maps representing the selected \textsc{fargo} phases shown in Fig.\,\ref{fig:4}. The two colors (greyshades) indicate, that the flux of the adjoint ring segment is set to zero and \textsc{AcDc} will compute a downward interpolation. For the subsequent spectrum calculations the observer is meant to look from the lower end of the box towards the central object.}%
    \label{fig:6}%
  \end{minipage}%
\end{figure}

For the calculation of the disk spectra we used the \emph{Ac}cretion \emph{D}isk \emph{c}ode (\textsc{AcDc}) developed by \citet{2004A&A...428..109N}. Starting with the assumption of a thin $\alpha$-disk as described by \citet{1973A&A....24..337S}, which allows us to decouple the vertical and radial structure, the disk is separated into concentric annuli. For each of these rings with radius $R$ the effective temperature $T_{\mathrm{eff}}$ as well as a viscosity times surface mass density $w\Sigma$ value can be derived for given white dwarf radius $R_{\mathrm{WD}}$, mass $M_{\mathrm{WD}}$, accretion rate $\dot{M}$, the gravitational constant $G$, and the Stefan-Boltzmann constant $\sigma$ from
\begin{eqnarray}
  T_{\mathrm{eff}}\left(R\right)&=&\left[\frac{3GM_{\mathrm{WD}}\dot{M}}{8\pi\sigma R^{3}}\left(1-\sqrt{\frac{R_{\mathrm{WD}}}{R}}\right)\right]^{\frac{1}{4}}%
  \label{eq:1}\\
  w\Sigma\left(R\right)&=&\frac{\dot{M}}{3\pi}\left(1-\sqrt{\frac{R_{\mathrm{WD}}}{R}}\right)\mathrm{.}%
  \label{eq:2}
\end{eqnarray}
The program then simultaneously solves the following set of equations for each of the rings:
\begin{itemize}
\item Radiation transfer for the specific intensity $I\left(\nu,\mu\right)$,
\item Hydrostatic equilibrium between gravitation, gas and radiation pressure,
\item Energy conservation for viscously generated $E_{\mathrm{mech}}$ and radiative $E_{\mathrm{rad}}$ and
\item Static rate equations $\frac{\partial n_i}{\partial t}=0$ for the population numbers $n_i$ of the atomic level $i$ of the given model atom in the non-LTE case.
\end{itemize}
The actual modification of the code took place in the subsequent spectral surface integration phase of \textsc{AcDc}. We constructed simple geometry maps (Fig.\,\ref{fig:6}) following the selected \textsc{fargo} pictures and reset the spectral flux for those ring segments, which are not part of the structure to zero. \textsc{AcDc} then integrates the whole disk spectrum by simple linear interpolation between the neighboring annuli spectra.
\section{Synthetic Spectra and Results}
\begin{figure}%
  \begin{minipage}{0.51\textwidth}%
    \centering%
    \includegraphics[width=\textwidth]{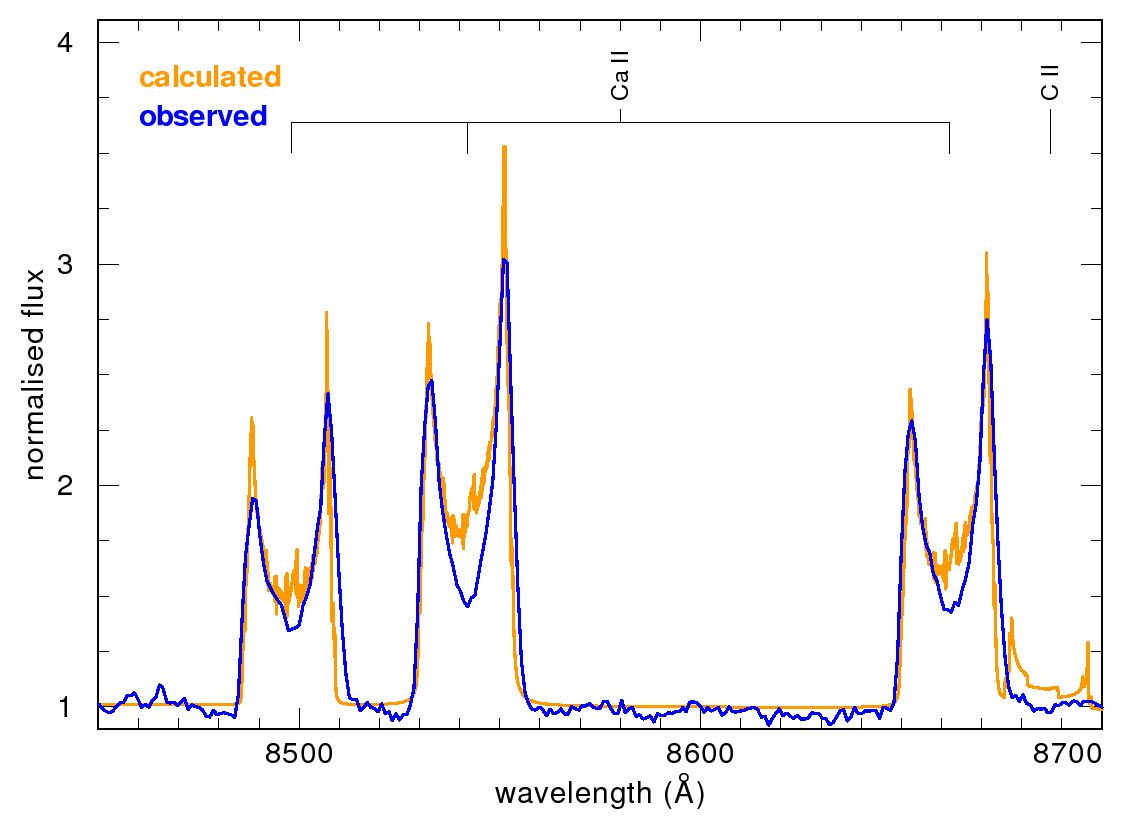}%
    \label{fig:7}%
  \end{minipage}%
  \hfill%
  \begin{minipage}{0.51\textwidth}%
    \centering%
    \includegraphics[width=\textwidth]{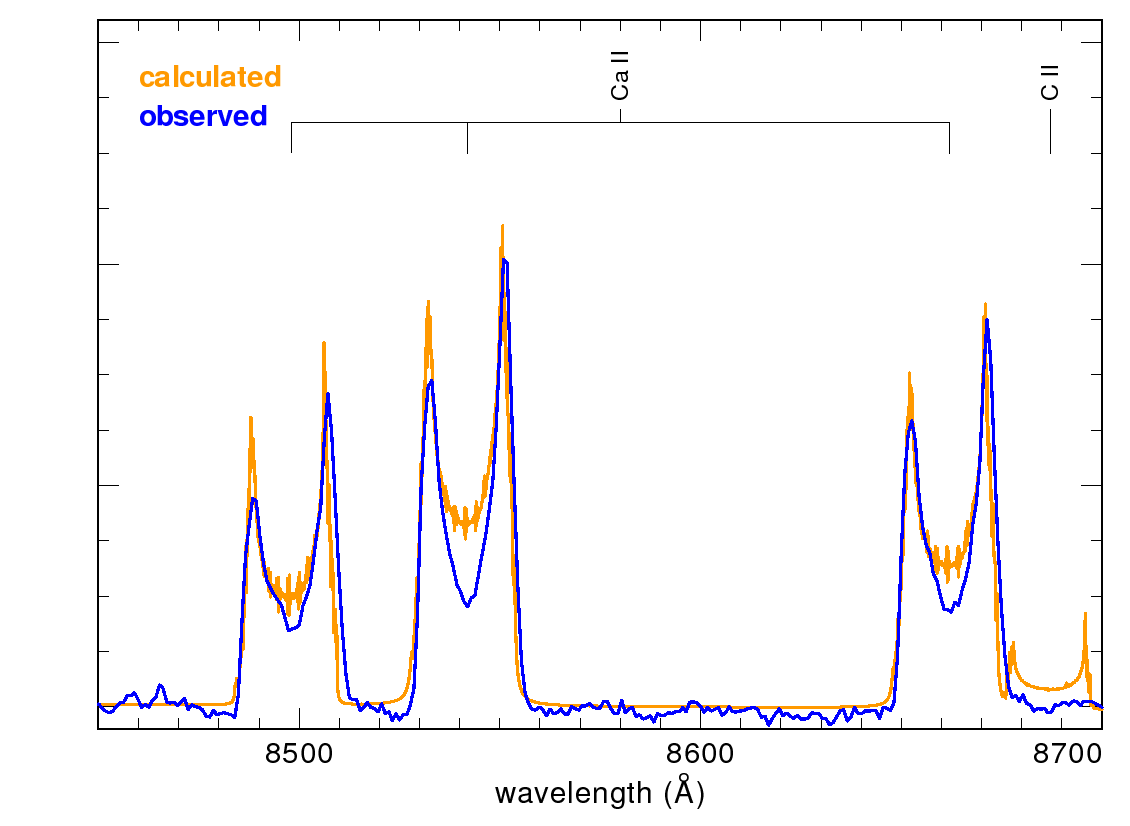}%
    \caption{Synthetic spectra for both geometries (shown in Fig.\,\ref{fig:6}), in comparison with the observation. The line asymmetry as well as the flat continuum between the lines are fitted well for an inclination of $i={76}^{\circ}$. The marked C\,\textsc{ii} feature seen to the far right of the modeled spectrum is actually a yet unsplitted doublet line.}%
    \label{fig:8}%
  \end{minipage}%
\end{figure}

Calculated with the set of parameters given in Tab.\,\ref{tab:2} and the model atoms (Tab.\,\ref{tab:3}), the spectra of the two geometry maps are shown in Fig.\,\ref{fig:8}.\\
\begin{table}%
  \centering%
  \begin{tabular}{llr}\hline
    Ring radii range&$R/R_{\mathrm{WD}}$&$2\,-\,136$\\
    Disk surface mass density&$\Sigma/(\mathrm{g}/{\mathrm{cm}}^{2})$&$0.3$\\
    Disk temperature range&$T_{\mathrm{eff}}/\mathrm{K}$&$6700\,-\,5600$\\\hline
  \end{tabular}%
  \caption{\textsc{AcDc} integration parameters.}%
  \label{tab:2}%
\end{table}%
\begin{table}%
  \centering%
  \begin{tabular}{l*{6}{r}}\hline
    Ions&H\,{\sc i-ii}& C\,{\sc i-iv}& O\,{\sc ii-iv}& Mg\,{\sc i-iii}& Si\,{\sc i-iv}& Ca\,{\sc i-iv}\\
    NLTE levels&$11$&$45$&$32$&$57$&$40$&$38$\\
    Line transitions&$45$&$88$&$65$&$104$&$49$&$64$\\
    Abundance (\%, in mass frac.)&$10^{-8}$&$4.6$&$65.5$&$13.5$&$15.1$&$1.3$\\\hline
  \end{tabular}%
  \caption{Model atoms and element abundances.}%
  \label{tab:3}%
\end{table}
The first result of the analysis is, that the inner radius of the disk must be increased to at least $R_{\mathrm{i}}{>}58\,R_{\mathrm{WD}}$ in order to suppress broad line wings and to fit the steep line profiles as well as the flat continuum between the triplet components. This value is slightly larger than the latest results by \citet{2010arXiv1007.2023M} with $R_{\mathrm{i}}=40\pm3\,R_{\mathrm{WD}}$.\\
The main goal of our work, matching the asymmetric line profile, also was achieved well for both geometries, although the relative strengths of the red/violet parts of each line seem to be generally estimated too low in the case of the closed disk structure.\\
One has also to keep in mind, that the asymmetry strongly depends on the orientation of the thicker region towards the actual line of sight, so the line profile should vary with the rotation of the asymmetrical disk around the central object. Even though \citet[priv.~comm.]{2008MNRAS.391L.103G} reported on a change in the line profile in the order of years for at least two of the known five objects, the predicted variability timescale of our model seems to be much shorter (order of hours) than the observations do suggest.\\
On the other hand, the \textsc{fargo} simulations suggest that there might also be a long-time variability as the disk evolves through different geometries in the time of hundreds of orbits. 
%
%
\begin{theacknowledgments}
  We thank Boris G\"ansicke for sending us his SDSS\,J1228+1040 spectrum in electronic form and for useful discussions. Also we like to thank Tobias M\"uller from the Computational Physics group of the University of T\"ubingen for providing us with the \textsc{fargo} code and the hydrodynamic simulations. T.R. is supported by the German Aerospace Center (DLR) under grant 05\,OR\,0806.
\end{theacknowledgments}
\bibliographystyle{aipproc}
\bibliography{shartmann}
\end{document}